# Starting to move through a granular medium

D. J. Costantino, T. J. Scheidemantel, M. B. Stone[†], C. Conger, K. Klein, M. Lohr, Z. Modig, P. Schiffer*

Department of Physics and Materials Research Institute, Pennsylvania State University, University Park, Pennsylvania 16802 USA

## Abstract

We explore the process of initiating motion through a granular medium by measuring the force required to push a flat circular plate upward from underneath the medium. In contrast to previous measurements of the drag and penetration forces, which were conducted during steady state motion, the initiation force has a robust dependence on the diameter of the grains in the medium. We attribute this dependence to the requirement for local dilation of the grains around the circumference of the plate, as evidenced by an observed linear dependence of the initiation force on the plate diameter.

*schiffer@phys.psu.edu

[†]Current address: Neutron Scattering Sciences Division, Oak Ridge National Laboratory, Oak Ridge, Tennessee, 37831, USA

Granular materials, collections of solid particles interacting only through contact forces, have a wide range of fascinating physical properties which are the subject of extensive current research efforts [1]. An important and intriguing property of granular materials is their response to externally applied stress, which differs considerably from that of homogeneous solids and fluids. When stressed, dense assemblies of these materials “jam” [2] by forming a rigid grain network which resists reorganization. This jamming influences granular drag, the force a dense granular medium exerts on a solid object which is pushed through the medium. Several groups have studied this process by measuring the drag force on objects moving horizontally through grains [3,4] or on vertical penetrators [5,6,7] as well as by imaging shearing processes [8]. This problem is also closely related to issues affecting the stability of structures embedded within the ground [9,10]. The results of these intruder experiments can be largely understood through continuum theories in which the exact size of the grains is not relevant, although large fluctuations in the force do reflect the granular nature of the medium [4].

In the present study, we measure the force needed to initiate upward vertical motion of a plate buried beneath spherical grains. In sharp contrast to previous measurements of the granular drag force, which were conducted in a steady state of motion, our measured initiation force has a substantial dependence on the diameter of the grains. This finding is a direct result of the granularity of the medium, and can be understood as a consequence of the need to move a volume of grains around the perimeter of the plate in order to initiate motion.

Our experimental apparatus (shown schematically in Fig. 1) measured the force needed to push a flat circular plate upward from the bottom of a cylindrical aluminum bucket of grains. The inner diameter of the bucket was 29.25 cm; measurements in smaller buckets indicated that the finite bucket diameter did not affect the measured force to within the uncertainty of the data (see supplementary information). The measured coefficient of static friction between glass beads and the aluminum was 0.18. Before filling the bucket, the plate (with diameter of $D_{plate}$) was centered and flush with the bucket bottom. We filled the bucket from a bead reservoir using a telescoping pipe approximately 40 cm long and of inner diameter 5.7 cm. This pipe was filled with grains while in contact with the container's base but off center. The pipe was then centered above the plate while still on the container's base and slowly retracted upwards to fill the bucket, with the grains never in freefall during filling as in the "localized-source procedure" described in reference [11]. This technique resulted in a consistent packing fraction of $\varphi = 59 \pm 1$ %. Pulling a sieve through the pile before taking data or filling from the container edge did not change the qualitative results, indicating that the results are not specific to the filling method (see supplementary information). All data shown are for container filling to a depth of $h_{pile}$ = ~15 cm, with the top of the pile brushed to be horizontal, but the results did not change qualitatively for $h_{pile}$ = 10 or 20 cm. A linear stepper motor pushed upward with constant velocity 0.025 cm/s on a spring ($k_{spring}$ = 3.4 N/cm for the smallest plate and 28 N/cm for the larger plates) which pushed a shaft attached to the intruder plate's bottom. The shaft entered the bucket through a low friction bearing, and a LVDT transducer recorded the motion of the plate. Tests demonstrated that the results were independent of the spring constant and velocity by comparing data acquired with the two springs and motor speeds 0.0053, 0.025, and 0.13 cm/s for a plate with $D_{plate}$ = 5.08 cm. We used a load cell attached to the stepper motor to record the force applied to the plate at a recording rate of 10 points per second. The grains were spherical glass beads (glass density of $\rho_g$ = 2.55 g/cm$^3$) with diameter, $d_{grain}$ = 0.054 ± 0.005, 0.08 ± 0.01, 0.15 ± 0.01, 0.20 ± 0.01, 0.22 ± 0.01, 0.303 ± 0.006, 0.494 ± 0.008 cm [12]. The plate diameter ranged from 1.27 cm to 10.16

cm. The humidity was 20.9 ± 0.7% for most runs, but humidity of 30-40% did not affect the observed trends in test measurements.

Fig. 2 shows data from two typical runs with different grain diameters and one plate diameter. Part (a) shows the force applied to the plate versus its displacement; a constant background force associated with the weight of the plate, shaft, and the armature of the transducer (1.09 – 1.81 N depending on the plate size) and friction between the shaft and the bearing (0.49 ± 0.01 N) is removed. Part (b) shows the plate's displacement versus time for these same runs. The insets of these plots show the complete data runs; the large drop in the force is associated with the grains being able to fit beneath the plate. This creates an artifact of an extremely large, grain-size dependent force from which the system does not relax even as the plate moves in the steady state.

In Fig. 2(a), we have used arrows to indicate the force, $F_{ini}$, needed to initiate the plate's motion. The region of data around $F_{ini}$ in Fig. 2 is clearly separable into two regimes: 1) when the plate is moving through the grains (right side of the figure) and the position is monotonically increasing; and 2) before the plate begins to move, where the data are dominated by electronic noise. We define $F_{ini}$ to correspond to the maximum in the curvature of $z(t)$, i.e. where its second derivative is maximum. Changing the parameters used to determine this point changes $F_{ini}$ by at most ~20 %.

From Fig. 2, we see that the $F_{ini}$ is clearly greater for the larger beads. This is quite surprising since previous drag and penetration data taken in steady state motion did not find any significant grain size dependence [4,5,6], and a particle size is not anticipated by continuum theories, e.g. as presented in [13]. The grain size dependence of $F_{ini}$ is apparently linear, as seen in Fig. 3 for a range of plate diameters. We obtain each point in the figure by averaging $F_{ini}$ from three to seven individual runs; the error bars are the standard deviations of this average.

To understand the origins of the grain size dependence of the initiation force, we first examine the dependence of $F_{ini}$ on the plate diameter, $D_{plate}$, as shown in Fig. 4 for representative values of $d_{grain}$. The inset shows that while $F_{ini}$ increases with $D_{plate}$, the dependence is not a simple quadratic relation $F_{ini} \sim D_{plate}{}^2$, which might be expected from hydrostatic pressure arguments. We are much better able to fit $F_{ini}(D_{plate})$ by using a functional form of $F_{ini}(D_{plate}) = AD_{plate} + BD_{plate}^2$. The main plot in Fig. 4 shows the data with the quadratic term of this fit subtracted, graphed against $D_{plate}$. The linear behavior of this graph confirms that our function, correctly describes the initiation force's dependence on the plate's diameter.

The grain-size dependence of the linear and quadratic coefficients, shown in Fig. 5, strongly suggests a physical basis for both quantities. The coefficient of the quadratic term, $B$, is essentially independent of grain size ($0.15 \pm 0.02$ N/cm$^2$). This is consistent with this term originating from the hydrostatic pressure associated with the weight of the grains, a continuum effect. A simple calculation based on the hydrostatic pressure at the container's base gives $B = (\varphi\rho_g g h_{pile}(\pi/4)) = (0.17 \pm 0.01$ N/cm$^2$), close to the experimental value we found, lending credence to this explanation. Different fillings of the bucket ($h_{pile}$), show a linear dependence of $F_{ini}$ with a slope consistent with hydrostatic pressure. The grain diameter independence of the quadratic term implies that the grain diameter dependence of $F_{ini}$ is associated with the linear term. This is substantially verified by finding a linear relation between $A$ and $d_{grain}$ in Fig. 5, mirroring the relation between $F_{ini}$ and $d_{grain}$

While we can easily explain the term $BD_{plate}^2$ in terms of a continuum model of granular media, the linear grain diameter dependence must be associated with the *granularity* of the material. As mentioned above other studies of steady state motion in spherical grains [4,5,6] found no grain size dependence on force. Reference [5] examined both downward and upward motion and reported that their results were particle size independent, implying that the upward direction of motion in our study is not responsible for the grain size dependence. A grain size dependence reported in reference [14] is attributed to differences in grain morphology in their angular particles, rather than originating explicitly in the grain size. Our finding of a robust grain size dependence is quite different, and it presumably is due to our measuring the force needed to initiate motion rather than the resistance force opposing a probe already in motion.

The difference between our strong grain-size dependence and the absence of such dependence observed in previous steady state resistance force measurements is rather striking. To develop an understanding of this result, we consider that a solid object pushing slowly through the grains will push a plug of grains ahead of its motion, as demonstrated explicitly in reference [10]. We propose that the grain size dependence is a result of *forming* this plug and separating it from the surrounding material. We note that the surface area of the side of the plug should be proportional to the plate's radius bolstering the notion that interactions along the sides are responsible for the linear term.

To develop a more detailed explanation for the observed grain size dependence associated with the linear term $AD_{plate}$, we consider the effect of varying a single parameter, $d_{grain}$, while holding the other relevant parameter, $D_{plate}$, constant. We hypothesize that the linear term is related to grain motion along the edges of the plug. For the plate to move, not only do the grains directly above the plate need to be lifted, interlocking grains around its perimeter need to be pushed aside. Dilation is thus needed around the perimeter of the plug, which requires the grains along the perimeter to be compressed into the bulk of the grains on either side to create a shear or "failure" zone about the sides of the plug. The affected volume of grains in the failure zone is a thin shell surrounding the plug, conceptually similar to the annulus of forbidden area which results in the grain size dependence in the Beverloo equation for granular flux through a hole [15]. This volume should be proportional to both the grain diameter and plate diameter, since the grains in the failure zone need to be moved aside by some fraction of their diameter, defining the thickness of the failure zone, and the circumference of the failure zone will be proportional to the plate's circumference. Different grain sizes will have the same coefficient of friction, number of contacts per grain, and confining pressure. Thus the resulting resistance to pushing aside the grains around the edge should be proportional to grain diameter and plate diameter in qualitative agreement with our findings. The grain size dependence of *A* in Fig. 5 can be fit to the linear form, $A(d_{grain}) = (0.03 \pm 0.06 \text{ N}) + (1.8 \pm 0.3 \text{ N/cm}^2)\, d_{grain}$. Note that the fitted slope is an order of magnitude larger than *B*. The force to compress grains into the bulk should be considerably larger than the force required to lift the grains above the plate, so this adds further credence to the above explanation.

Our results demonstrate a new manifestation of how the *granularity* of granular materials, specifically the grain size, can affect even bulk properties associated with a macroscopic number of grains. Furthermore, the results demonstrate an important phenomenon; initiating motion through the grains appears to be distinctly different from continuous motion. The data also have practical implications for the motion of objects through granular materials, for example in cases where objects buried in grains should be held stable (e.g., a post mounted in sand) or need to begin motion (e.g., halted mixer blades). Future local experimental studies, through MRI or confocal microscopy, or detailed computational modeling of the process could shed additional insight into the grain-scale motion responsible for these results.

We acknowledge NASA funding through grant NAG3-2384 and NSF for assistance through the REU program. We are also grateful for helpful conversations with Jayanth Banavar and Robert Behringer.

Figures

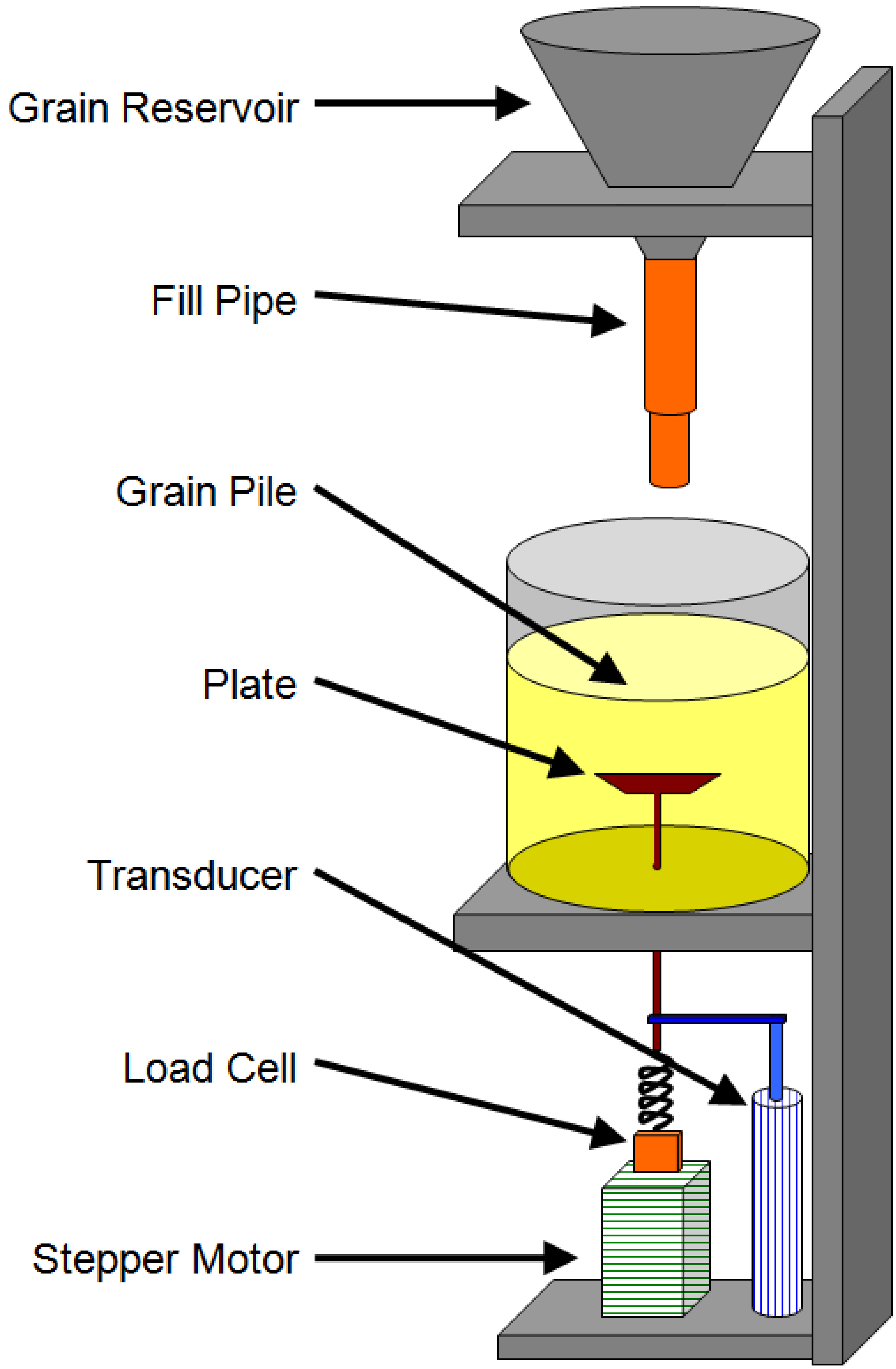

FIG. 1

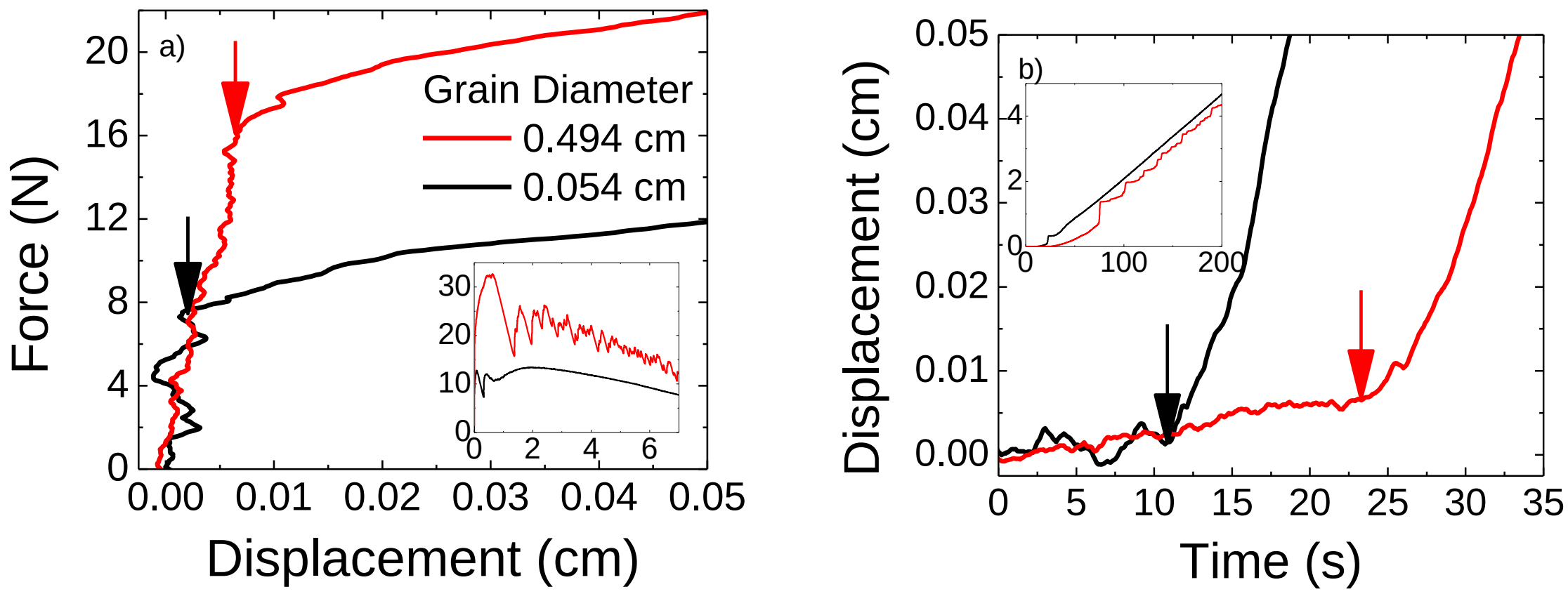
a)
Force (N)
Displacement (cm)
Grain Diameter
0.494 cm
0.054 cm
b)
Displacement (cm)
Time (s)

FIG. 2

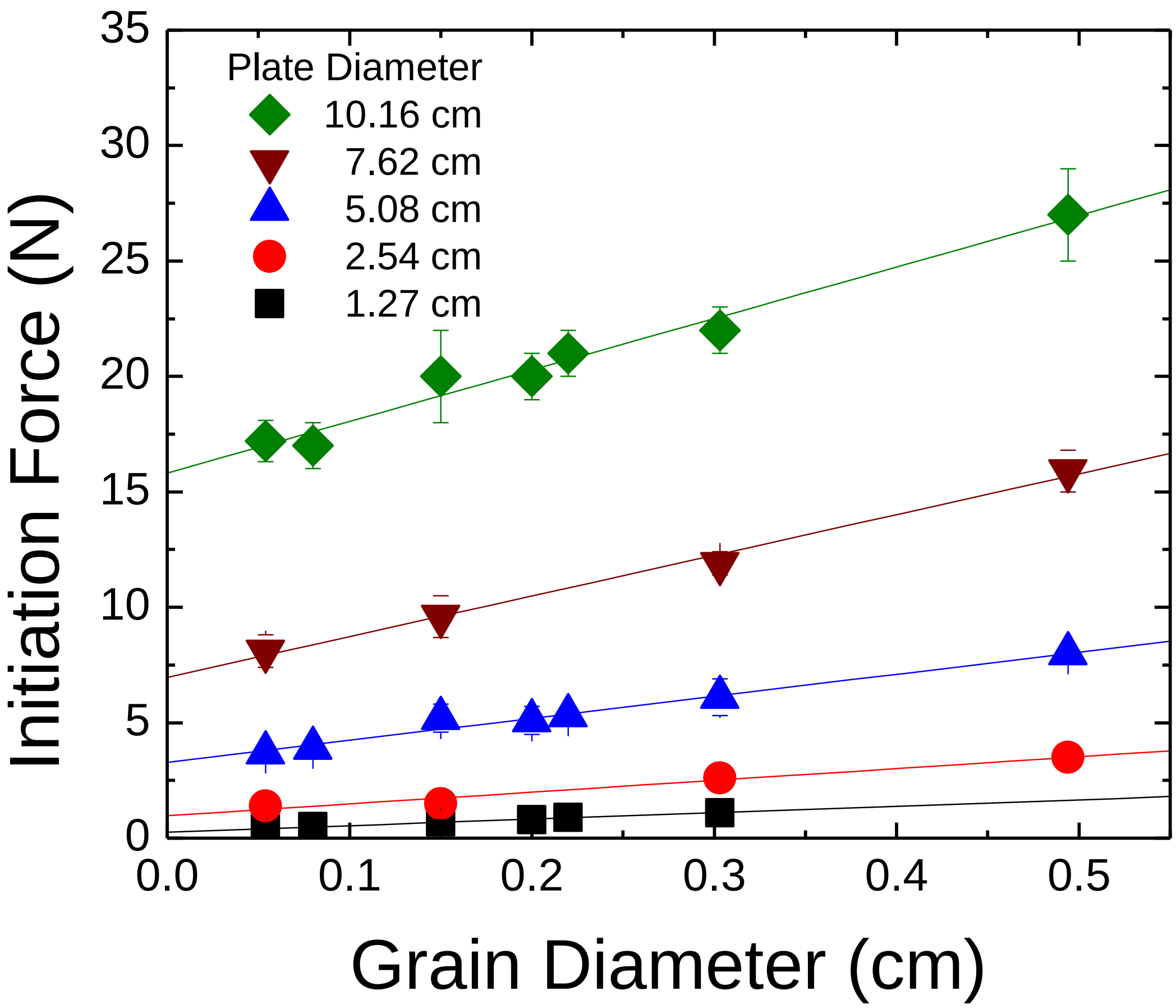
Plate Diameter
10.16 cm
7.62 cm
5.08 cm
2.54 cm
1.27 cm
Initiation Force (N)
Grain Diameter (cm)
0
5
10
15
20
25
30
35
0.0
0.1
0.2
0.3
0.4
0.5

FIG. 3

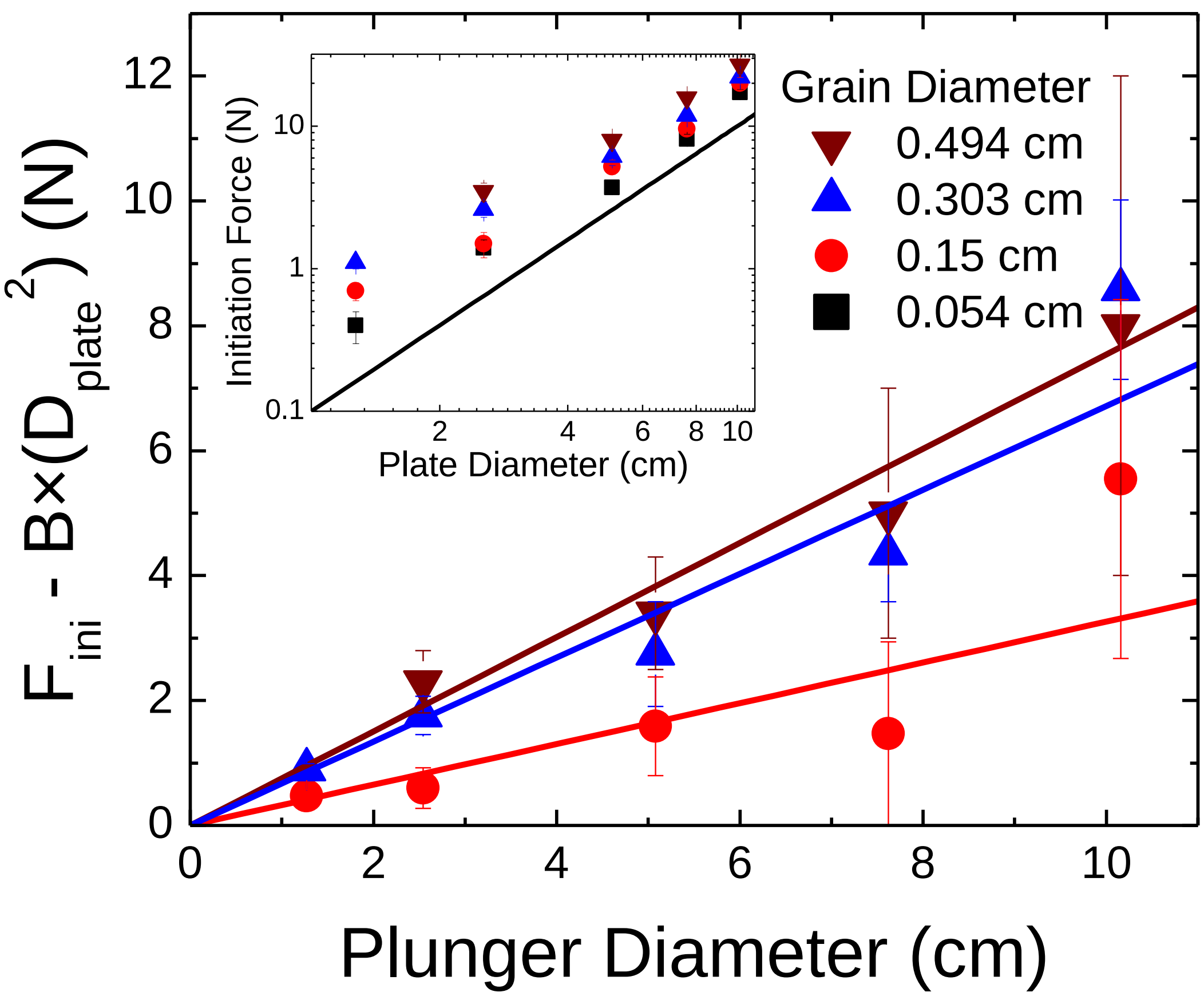
Grain Diameter
0.494 cm
0.303 cm
0.15 cm
0.054 cm
Initiation Force (N)
Plate Diameter (cm)
$F_{ini} - B\times(D_{plate}^2)$ (N)
Plunger Diameter (cm)

FIG. 4

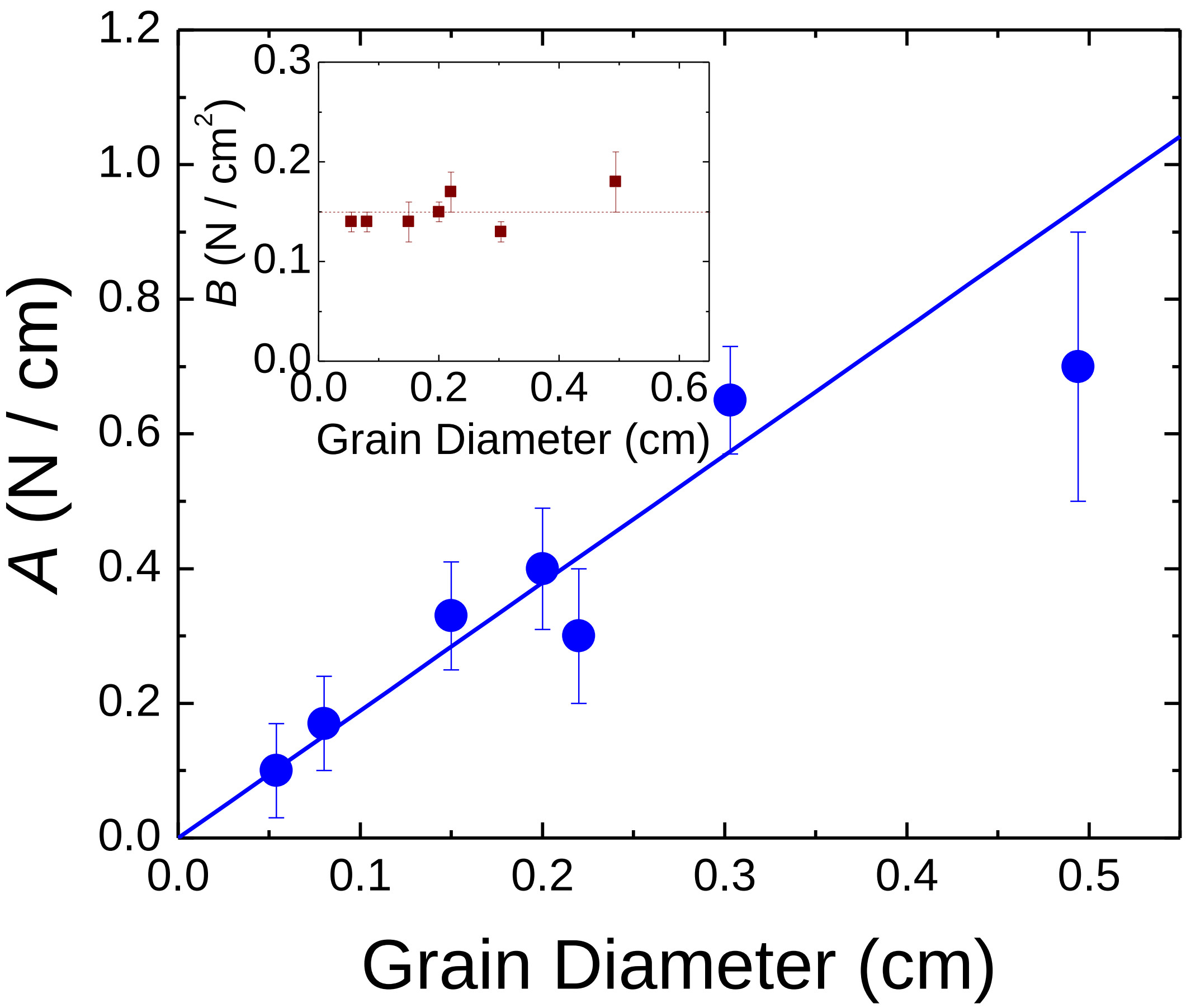

A (N / cm)
Grain Diameter (cm)
0.0
0.2
0.4
0.6
0.8
1.0
1.2
0.0
0.1
0.2
0.3
0.4
0.5
B (N / cm²)
0.0
0.1
0.2
0.3
0.0
0.2
0.4
0.6
Grain Diameter (cm)

FIG. 5

**Figure Captions**

**FIG. 1** Schematic drawing of our experiment as described in the text.

**FIG. 2** Typical raw data from our measurements. The plate diameter was 7.62 cm, and the $d_{grain}$ = 0.494 cm and 0.054 cm for the red and black curves respectively. a) Force vs. plate displacement near the point where motion is initiated; the weight and friction forces from the bearing have been subtracted from the data. The initiation force, $F_{ini}$, is indicated by arrows for both runs and is clearly larger for the larger diameter beads. b) Data for the same runs showing the plate's displacement vs. time; the times at which $F_{ini}$ was reached are indicated by arrows. Both insets show broader ranges of data.

**FIG. 3** The grain size dependence of the force required to initiate motion, $F_{ini}$, for different plate diameters. The lines are linear fits to the data.

**FIG. 4** The linear term of the initiation force's plate size dependence for three different grain diameters. The points are generated by subtracting the quadratic term from the fits of $F_{ini} = AD_{plate} + BD_{plate}^2$, in which we weight the data points by the uncertainty for each datum. The linear behavior of $(F_{ini} - BD_{plate}^2)$ confirms the fits' validity. The log-log plot in the inset shows $F_{ini}$ vs. $D_{plate}$; the dashed line has slope of 2; the data cannot be described by a simple quadratic form. The smallest grain diameter was not included in the main plot because the quantity $AD_{plate}$ is indistinguishable from zero within experimental uncertainty for this grain size.

**FIG. 5** Linear coefficient from $F_{ini} = AD_{plate} + BD_{plate}^2$ plotted against $d_{grain}$. The solid line, guide for the eye, is a linear fit to the data after weighting the data by each point's error and forced to pass through the origin. The inset gives the quadratic coefficient, $B$ as a function of $d_{grain}$. The dashed line in the inset is the average value of this coefficient.